\documentclass[fleqn,usenatbib]{mnras}

\usepackage{newtxtext,newtxmath}
\providecommand{\apspr}{Astrophys.~Space~Phys.~Res.}  % added by Yilong

\usepackage[T1]{fontenc}

\DeclareRobustCommand{\VAN}[3]{#2}
\let\VANthebibliography\thebibliography
\def\thebibliography{\DeclareRobustCommand{\VAN}[3]{##3}\VANthebibliography}

\usepackage{graphicx}	% Including figure files
\usepackage{amsmath}	% Advanced maths commands

\hypersetup{colorlinks=true, urlcolor=blue}
\usepackage{xcolor}     % added by ylwang for colored text
\usepackage[normalem]{ulem} % added by ylwang to mark deleted text

\defcitealias{wang2024}{Paper I}

\title[Accretion with two-phase gas supply]{Accretion with two-phase gas supply and its application in black hole X-ray binaries}

\author[Wang et al.]{
Yilong Wang,$^{1,2,3,4}$\thanks{wangyilong@nao.cas.cn} 
B. F. Liu,$^{1,2}$\thanks{bfliu@nao.cas.cn}
and Mingjun Liu$^{1}$
\\
$^{1}$National Astronomical Observatories, Chinese Academy of Sciences, 20A Datun Road, Beijing 100101, China\\
$^{2}$School of Astronomy and Space Science, University of Chinese Academy of Sciences, 19A Yuquan Road, Beijing 100049, China\\
$^{3}$Institute of Space Sciences (ICE, CSIC), Campus UAB, Carrer de Can Magrans s/n, 08193, Barcelona, Spain\\
$^{4}$Institut d’Estudis Espacials de Catalunya (IEEC), Carrer Gran Capità 2–4, 08034, Barcelona, Spain}

\date{Accepted XXX. Received YYY; in original form ZZZ}

\pubyear{\the\year{}}

\begin{document}
\label{firstpage}
\pagerange{\pageref{firstpage}--\pageref{lastpage}}
\maketitle

% Abstract of the paper
\begin{abstract}
Accretion in black hole X-ray binaries is commonly believed to be supplied by the Roche lobe overflow or the stellar wind. The former is thought to form a geometrically thin disc while the diffuse wind could form a geometrically thick hot accretion flow. In this paper, we instead consider a more generalised case, i.e., accretion with both cold and hot gas supplies, which feed a disc and a corona respectively. We investigate the interaction of disc and corona by analysing the energy coupling and matter exchange, i.e. corona condensation/disc evaporation, with a semi-analytical method. It is found that the accretion geometry in the radial direction and the resultant emission spectrum depend strongly on both the total gas supply rate and the ratio of cold and hot gases. For gas supply rates of a few percent of the Eddington value, diverse geometries and spectral shapes are possible, depending on the fraction of cold gas supply. This provides an interpretation for the various spectra observed in intermediate states. However, at higher accretion rates, regardless of the form of the feeding gas, the inner accretion flow is always disc-dominated, implying an inevitable transition to the soft state, while at very low gas supply rates, hard state spectrum dominated by the hot flow is expected. We also present the predicted hardness-intensity correlation of Cygnus X-1, and constrain the value of the viscosity parameter of the accretion flow to the range of $0.25$--$0.35$ by comparing our results with MAXI observations. 
\end{abstract}
% 246 words in the abstract

% Select between one and six entries from the list of approved keywords.
% Don't make up new ones.

%%%%%%%%%%%%%%%%%%%%%%%%%%%%%%%%%%%%%%%%%%%%%%%%%%

%%%%%%%%%%%%%%%%% BODY OF PAPER %%%%%%%%%%%%%%%%%%

\section{Introduction} \label{sec:intro}
Black hole X-ray binaries (BHXRBs) are observed to exhibit different spectral states. Among them, the most well-studied are the high/soft state, the low/hard state and the intermediate state \citep[for reviews, see][]{remillard2006, done2007}. The high/soft state spectrum in X-ray band is generally believed to originate from the multicolour black body radiation of a standard thin disc \citep[][]{shakura1973,novikov1973} which extends to the innermost stable circular orbit (ISCO) of the black hole. On the other hand, in the low/hard and intermediate states, a hard X-ray power-law component appears and often dominates the spectrum. This power-law component is most likely produced by inverse Compton scatterings in a hot, diffuse medium around the black hole, which is either an advection-dominated accretion flow (ADAF) or a corona lying above a disc, or even an outflow in some special cases. The hot ADAF/corona and the cold disc together form the radial or vertical two phases of the accretion flow, which has successfully explained the most prominent observed spectral characteristics of BHXRBs (and also active galactic nuclei, AGNs). However, the exact geometry of accretion flows near black holes remains unclear. Debates arise from X-ray variability and reverberation analyses, which point to diverse geometries of two-phase accretion flows \citep[e.g.][]{done1989, fabian2009, demarco2011, kara2013, reis2013, cackett2014, emmanoulopoulos2014, uttley2014, poutanen2018, karpouzas2021}{}. 

A common characteristic deduced from observational diagnostics is that the hard X-ray radiation region is very small, located within a few tens of gravitational radii around the black hole \citep[e.g.][]{reis2013, fabian2015, liu2017}{}. Thus, a compact corona as the source of the power-law X-ray spectrum is widely adopted \citep[e.g.][]{risaliti2011, sanfrutos2013, wilkins2011, wilkins2012, wilkins2016}{}. On the other hand, specific features appear in some individual sources, such as the varying fraction of reflection component in hard X-rays, leading to a speculation of the hot flow being outflow or jet base \citep[e.g.][]{beloborodov1999, malzac2001, markoff2001, merloni2002, markoff2005, you2021, wangjy2022}. Nevertheless, recent studies of the observed X-ray polarization in BHXRBs \citep[e.g.][]{krawczynski2022, veledina2023, ingram2024} support the idea that the corona lies in a plane parallel to the disc.

Theoretically, it is natural to accept that the hot gas is a part of the accretion flow, because the accretion flow not only continuously replenish hot gas, but it also emits X-rays dominantly in the innermost region \citep[e.g.][]{liu2017}{}, as expected for a small, compact corona. In general, such accretion flows are believed to be a mixture of cold and hot flows, either in a form of an inner hot flow connecting to an outer truncated accretion disc \citep[e.g.][]{narayan1996, narayan2008, dove1997, esinetal1997, esin1998, poutanen1997, liu1999, meyer2000b}, or a disc sandwiched between coronae \citep[e.g.][]{liang1977, haardt1991, haardt1993, svensson1994, liu2022}. There were also investigations of patchy corona covering part of the disc surface \citep[e.g.][]{galeev1979, haardt1994, stern1995, svensson1996} and outflowing corona \citep[e.g.][]{beloborodov1999, malzac2001, merloni2002}.

In the framework of disc-corona interaction, the theoretical investigations of two-phase accretion flows predict the variation of hot and cold flows with the changing accretion rate, thus providing a natural mechanism for the spectral transitions in black hole systems. \citep[][] {meyer1994, liufk1995, liu1999, meyer2000a, meyer2000b, rozanska2000a, rozanska2000b, liu2002, meyer-hofmeister2003, meyer-hofmeister2005, liu2006, liu2007, meyer2007, taam2008, liu2011, qiao2012, qiao2013a, qiao2018b, taam2018, meyer-hofmeister2020, cho2022, wang2024}. For a review, see \citet[][]{liu2022}. In these works, the gravitationally captured gas was assumed to either feed the disc as in Roche-lobe overflow accretion systems or directly supply the hot accretion flow as in wind accretion systems, and the processes of disc evaporation or hot gas condensation were  investigated respectively.

In this work we consider a more realistic environment around black holes where hot and cold gas supplies co-exist. Following the earlier investigations, we study the energy coupling and matter exchange between the two phases of the accretion flows. The complexity of accretion with these two types of gas supplies means both evaporation and condensation processes could happen, and this work marks our first effort to unify corona condensation and disc evaporation in one single model. This is achieved on the basis of the improved corona condensation model described in \citet[][]{wang2024}, hereafter \citetalias{wang2024}. We generalize this model to allow different initial composition of the accretion flow and self-consistently include evaporation process in the calculation. More specifically, we assume that the accretion starts with a sandwich geometry where the accreted gas feeds both the corona and the disc. With different composition of two-phase gas supply, we investigate the change of the geometry of the accretion flow along distance as a consequence of disc-corona coupling. We then calculate the corresponding emergent spectrum by Monte Carlo simulations \citep[][]{pozdniakov1977, pozdnyakov1983, manmoto1997, qiao2012, ghosh2013}.

The aim of this study is to interpret the diversified geometries and spectra observed in BHXRBs (and potentially in AGNs). By assuming various possible compositions of gas supply to the accretion flows in either low mass X-ray binaries (LMXBs) or high mass X-ray binaries (HMXBs), we examine the final, stable geometries of the accretion flows in the neighbourhood of the black holes, which produce the majority of the observed radiations and therefore determine the spectra. We also investigate the two extreme cases where gas is supplied to a disc alone or to an ADAF alone at different accretion rates.

We pay special attention to accretion flows with intermediate gas supply rates. In this case, various geometries and X-ray spectra at a similar level of X-ray luminosity are expected as a result of different compositions of hot and cold gas supplies. We also analyse how the spectral states of BHXRBs are determined by the property of the accreted gas.

In Section \ref{sec:model} we summarise the basics of our model. In Section \ref{sec:results} we present the numerical results. In Section \ref{sec:discussion} we discuss the application of the theoretical results in observation. Our conclusion is presented in Section \ref{sec:conclusions}.

\section{The Model} \label{sec:model}
Here we briefly review the physics of condensation and evaporation, emphasizing the modifications we have made since \citetalias{wang2024}, and more details can be found in the Section 2 of \citetalias{wang2024}.

\subsection{Form of the accreted gas} \label{subsec:form of gas}
We consider an accretion flow around black hole fed by both cold and hot gases, which respectively supply an accretion disc and its corona. The initial status of the accretion flow can therefore be described by a two-phase flow specified by two parameters, the initial accretion rate of the disc, $\dot{m}_{\rm disc,ini}$, and that of the corona, $\dot{m}_{\rm cor,ini}$. Thus, the total accretion rate $\dot{m}$ is given by $\dot{m} =\dot{m}_{\rm disc,ini} +\dot{m}_{\rm cor,ini}$. In practice, equivalently we use $\dot{m}_{\rm disc,ini}$ and $\dot{m}$ as input parameters for numerical calculations. By adjusting $\dot{m}_{\rm disc,ini}$ for each given $\dot{m}$, we cover a wide range of possibilities of the accretion mode. The limit $\dot{m}=\dot{m}_{\rm disc,ini}$ corresponds to the well-studied thin disc accretion, whereas the limit of $\dot{m}=\dot{m}_{\rm cor,ini}$ corresponds to accretion via hot, diffuse stellar wind or interstellar medium, as studied by \citet{liu2015,taam2018,qiao2018b,meyer-hofmeister2020} and recently improved by \citetalias[][]{wang2024}. In this work we investigate the generalized situation. 

We study the interaction between the pre-existing disc and corona, taking into account the radiation coupling, thermal conduction and enthalpy flux between the two phases. Our aim is to investigate how such two-phase accretion flows evolve on the way of accreting to the black hole, and what the stable accretion geometry is, in particular, in the innermost region where most of the radiations are produced. By changing the total accretion rate and the ratio of cold and hot gas supplies, we are able to study how the geometry and spectrum vary with the accretion rate and the observed X-ray luminosity.

\subsection{Vertical stratification of the accretion flow} \label{subsec:vertical stratification}
We use a semi-analytical method to study the interaction between the two phases, corona and disc, of the accretion flow. To achieve this, we assume that the accretion flow is vertically stratified into three layers, the corona, the disc, and a transitional layer in between where matter exchange, i.e., corona condensation and disc evaporation, happens \citep[e.g.][]{liu2007,meyer2007}. 

The corona is similar to an ADAF \citep[][]{narayan1994,narayan1995a,narayan1995b} except that its physical properties are altered by its interaction with the underlying disc. Most importantly, the inverse Compton scattering of disc soft photons in the corona and the thermal conduction between corona and disc serve as extra cooling terms in the energy balance equation of the corona, which lower the electron temperature and increase the radiative efficiency of the corona.

The transitional layer is a cooler and denser thin layer of the corona which has coupled ion and electron temperatures \citep[see Fig. 3 in][]{meyer2007}, in contrast to the main part of the corona where the ion and electron temperatures are decoupled due to inefficient Coulomb collisions in low densities. In the transitional layer, the dominant heating mechanism is the thermal conduction from the corona, and cooling is dominated by bremsstrahlung. Since conductive heating and bremsstrahlung cooling do not necessarily cancel each other out, energy balance in the transitional layer has to be restored by the enthalpy carried by the condensed/evaporated gas. More specifically, when conductive heating exceeds cooling, a certain amount of cool gas is heated up into the corona, i.e., evaporated, which usually happens in the outer part of the accretion flow. The opposite could happen in the denser inner region if bremsstrahlung cooling exceeds conductive heating and the hot gas is over-cooled and therefore condenses into the disc. 

The disc we consider in our model is effectively a Shakura-Sunyaev disc \citep[][]{shakura1973} with accretion rate varying with radius, due to the condensation/evaporation effects. Moreover, we include the irradiation of the disc by the corona, where part of the coronal emission is reprocesses by the disc and consequently increases the disc temperature and its emission. We have also considered the non-local effect of inner disc illuminating the outer part of the corona, which provides an extra source of soft photons for Compton scattering, although this is not important in the inner region where photons from local disc emission (including the contribution of coronal irradiation) dominate.

\subsection{Modification on thermal conduction} \label{subsec:conduction}
As the dominant heating mechanism in the transitional layer, the thermal conduction between the two phases of the accretion flow has a significant influence on the condensation/evaporation processes. Weaker thermal conduction leads to inefficient heating in the transitional layer and hence over-cooling and effective condensation. For the same reasoning, stronger thermal conduction tends to suppress condensation and enhance evaporation. Moreover, thermal conduction also serves as the dominant cooling process in the outer part of the corona, where the density is low and therefore the radiative cooling processes are weak. The electron temperature profile of the corona thus can be significantly different from that of a pure ADAF due to thermal conduction. 

In our previous works (including \citetalias{wang2024}), the conductive flux was approximated as \citep[see also equation (6) in][]{qiao2013a},
\begin{equation}
    \Delta F_{\rm c} = k_0 T_{\rm e}^{\frac{5}{2}} \frac{dT_{\rm e}}{dz} \approx k_0 T_{\rm e}^{\frac{5}{2}} \frac{T_{\rm e} - T_{\rm cpl}}{H} \approx k_0 T_{\rm e}^{\frac{7}{2}}/H,
    \label{eq:old conduction}
\end{equation}
where $T_{\rm e}$ is the electron temperature in the corona, $T_{\rm cpl}$ is the coupling temperature in the transitional layer which is neglected here as compared to $T_{\rm e}$, $H$ is the scale height, and $k_0 = 10^{-6}{\rm \,erg\,s^{-1}\,cm^{-1}\,K^{-7/2}}$ \citep[][]{spitzer1962}.

We now replace the above approximation to a more precise formula, that is, 
\begin{equation}
     \Delta F_{\rm c} = k_0 T_{\rm e}^{\frac{5}{2}} \frac{dT_{\rm e}}{dz}  =\frac{2}{7} k_0 \frac {d T_{\rm e}^{\frac{7}{2}}}{dz} \approx \frac{2}{7} k_0 \frac{T_{\rm e}^{7/2} - T_{\rm cpl}^{7/2}}{H}.
    \label{eq:new conduction}   
\end{equation}
Note that the extra factor of $\frac{2}{7}$ and the inclusion of $T_{\rm cpl}$ give a smaller conductive flux, which leads to more effective condensation (or less effective evaporation) in this work as compared to \citetalias{wang2024} and previous relevant studies.

\subsection{Input parameters}\label{subsec:parameters}
The input parameters of our semi-analytical model include black hole mass ($m$) scaled with solar mass, accretion rate ($\dot{m}$) scaled with Eddington accretion rate ($\dot{M}_{\rm Edd} = 1.39 \times 10^{18} m$), viscosity parameter ($\alpha$), magnetic parameter ($\beta$) which is defined as the ratio of gas pressure to the sum of gas pressure and magnetic pressure, and the albedo ($a$) of coronal irradiation at disc surface \citepalias[see also][]{wang2024}. In addition, we assume that a part ($\dot{m}_{\rm disc,ini}$) of the total accretion rate ($\dot{m}$) supplies initially to a disc at a distance of $R_{\rm out}$, unlike in \citetalias[][]{wang2024} where $\dot{m}_{\rm disc,ini} = 0$ was implicitly assumed.

\section{Numerical results} \label{sec:results}
The main focus of this work is to investigate the changes of the accretion geometry and the emergent spectrum in response to variations in the amount and composition of the accreted gas, which are respectively represented by $\dot{m}$, the total accretion rate, and $\dot{m}_{\rm disc,ini}$, the initial accretion rate via the disc. Therefore, the numerical results presented in this work are derived with the other input parameters fixed to $m = 10$, $\alpha = 0.3$ \citep[][]{king2007}, $\beta = 0.95$ \citep[][]{hawley2001}, $a = 0.15$ \citep[][]{magdziarz1995} and $R_{\rm out} = 500 R_{\rm S}$ (here $R_{\rm S}$ is the Schwarzschild radius), unless stated otherwise. The effect of $R_{\rm out}$ is discussed later in Section \ref{subsec:rout} and the effect of $\alpha$ is demonstrated in Section \ref{subsec:hmxb}.  

In this work we confine our study to sub-Eddington accretion, with $\dot{m}$ ranging from $0$ to $1$. We consider all possible compositions of hot and cold gas supplies, thus the value of $\dot{m}_{\rm cor,ini}$ should, in principle, change from $0$ to $\dot{m}$. However, since a steady hot accretion flow exists only below a certain critical accretion rate (see Section \ref{subsec:constraint} for a detailed discussion), the results we present here are also constrained by this critical value. If the true supply rate of hot gas exceeds the upper limit, a part of the hot gas could clump/collapse into a disc by instability, with the remaining hot gas  forming a steady hot accretion flow at an accretion rate equal to or less than the critical value. The subscript "ini" in the symbols of the accretion rates denotes the values at $R_{\rm out}$, where we assume steady accretion flows are already formed.

\subsection{The maximal accretion rate in the hot flow as a signature of spectral state transition} \label{subsec:constraint}
The critical accretion rate of ADAFs has been discussed in the literature \citep[e.g.][]{narayan1995b, mahadevan1997, narayan1998, li2023, liu2025}, which is a few percent of the Eddington value (depending on the viscosity parameter). When the accretion rate exceeds this critical value, the Coulomb coupling between ions and electrons becomes so efficient that a two-temperature hot flow can no longer be preserved. Such a critical accretion rate also varies with distance. In general, the closer to the central black hole, the smaller it becomes. Naturally, when a cool disc underlies such a hot flow, this upper limit of the accretion rate in the hot corona is substantially affected by the interaction between the hot and cold flows, which are coupled by radiation and gas evaporation/condensation. For a given value of the total accretion rate ($\dot m$), we calculate the mass exchange and the structure of the two-phase accretion flows along the radial direction while gradually increasing the fraction of the hot gas at the outer radius ($R_{\rm out}$), to obtain the maximal accretion rate ($\dot{m}_{\rm cor,ini,max}$) allowed in the hot flow at $R_{\rm out}$. When $\dot{m}_{\rm cor,ini}>\dot{m}_{\rm cor,ini,max}$, no solutions can be found. Performing calculations for a series of total accretion rates ($\dot m$), we find that the maximal rate depends on $\dot{m}$.

\begin{figure}
\centering
\includegraphics[width=\columnwidth]{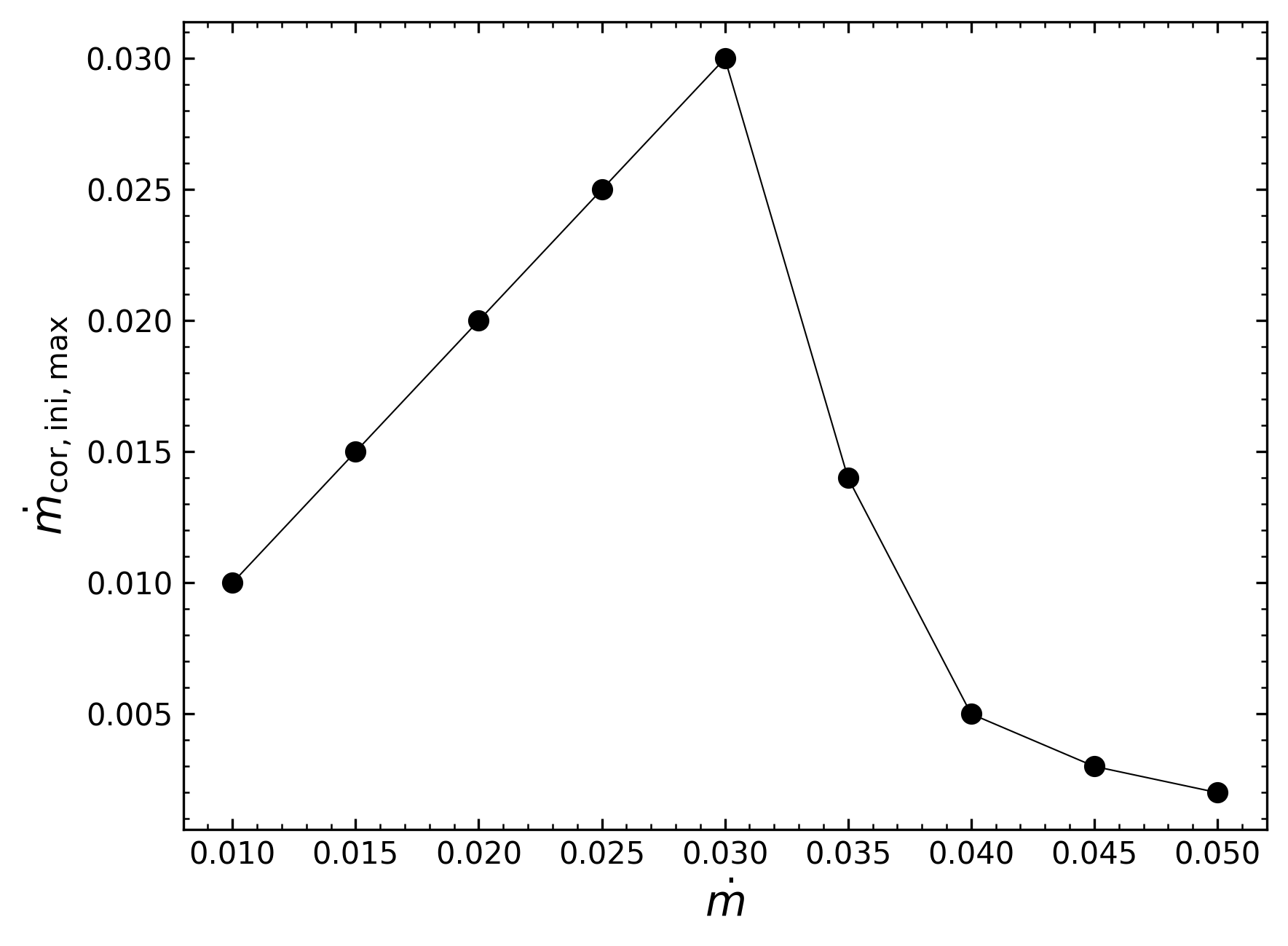}
\caption{The highest possible values of initial accretion rate in the corona, $\dot{m}_{\rm cor,ini,max}$, with $\dot{m}$ ranging from $0.01$ to $0.05$.}
\label{fig:mc_max}
\end{figure}

In Fig. \ref{fig:mc_max}, we plot the maximal value, $\dot{m}_{\rm cor,ini,max}$, as a function of the total accretion rate. It can be seen that for $\dot{m}$ up to $\sim 0.03$, it is possible for the accretion to be entirely via the hot flow, which corresponds to pure hot gas accretion presented in  \citetalias[][]{wang2024}. However, as $\dot{m}$ exceeds $\sim 0.03$, $\dot{m}_{\rm cor,ini,max}$ decreases with increasing $\dot{m}$, indicating that the supplied gas cannot all accrete via the corona, instead a fraction of it must accrete via a disc. The allowed gas flow in the corona is  smaller than $0.03$ because of the extra cooling caused by the underlying disc. The further increase of the total accretion rate leads to even more efficient cooling and rapid decrease of the allowed gas flow rate in the corona. If the supplied gas does have $\dot{m}_{\rm cor,ini} > \dot{m}_{\rm cor,ini,max}$, we expect that a fraction of the hot gas would soon collapse due to efficient cooling, and $\dot{m}_{\rm cor,ini} < \dot{m}_{\rm cor,ini,max}$ would be achieved by the time the accretion flow reaches equilibrium.

This dependence of $\dot{m}_{\rm cor,ini,max}$ on the total accretion rate implies a spectral state transition around $\dot{m} \sim 0.03$. For high supply rates, say, $\dot{m} \gtrsim 0.03$, the initial accretion is dominantly via a disc because the upper limit of accretion rate in the coronal flow is lower than a few percent of the Eddington value. When such two-phase gas accretes toward the black hole, the coronal gas can easily condense into the disc, as a consequence of strong Compton cooling. The accretion flow near the black hole is a thin disc together with a negligible corona, which does not really depend on the initial ratio of hot and cold gas supply rates at the outer boundary. Therefore in the high state the spectrum is generally characterised by disc emission, which is a multi-colour black body.

For very low gas supply rates, say, $\dot m<0.01$, it has been demonstrated \citep[e.g.][]{meyer2000b,liu2002,liu2022} that even with pure cold gas supply, the disc can be fully evaporated into an ADAF before it reaches $R_{\rm out} = 500R_{\rm S}$, the outer boundary we adopted in this work. On the other hand, in the case of pure hot gas supply, an ADAF will keep its form at all distances down to the ISCO \citepalias[e.g.][]{wang2024}. For the generalised situation of accretion with both hot and cold gas supplies, given such low accretion rates, even though there can be differences in the outer region, the accretion flow always converges to an ADAF in the inner region where most of the observed emission is produced. Therefore, for very low accretion rates, we observe a typical hard state spectrum dominated by the ADAF, irrespective of the composition of the initial gas supply.

For intermediate gas supply rates, $0.01 \la \dot{m} \la 0.03$, a wide range of initial accretion rate in the corona is possible. Thus, the accretion geometry is anticipated to depend strongly on both the total supply rate and its composition of hot and cold gases, which may interpret the diversity of the observed continuum and line emissions in BHXRBs with intermediate Eddington ratios ($L/L_{\rm Edd}$, where $L$ is the bolometric luminosity and $L_{\rm Edd} = 1.26 \times 10^{38}m$ is the critical Eddington luminosity).    

The trend of spectral state transition described above can be clearly seen from the hardness-intensity diagram (HID) calculated from the model.

\subsection{Hardness-intensity diagram} \label{subsec:HID}
\begin{figure}
\centering
\includegraphics[width=\columnwidth]{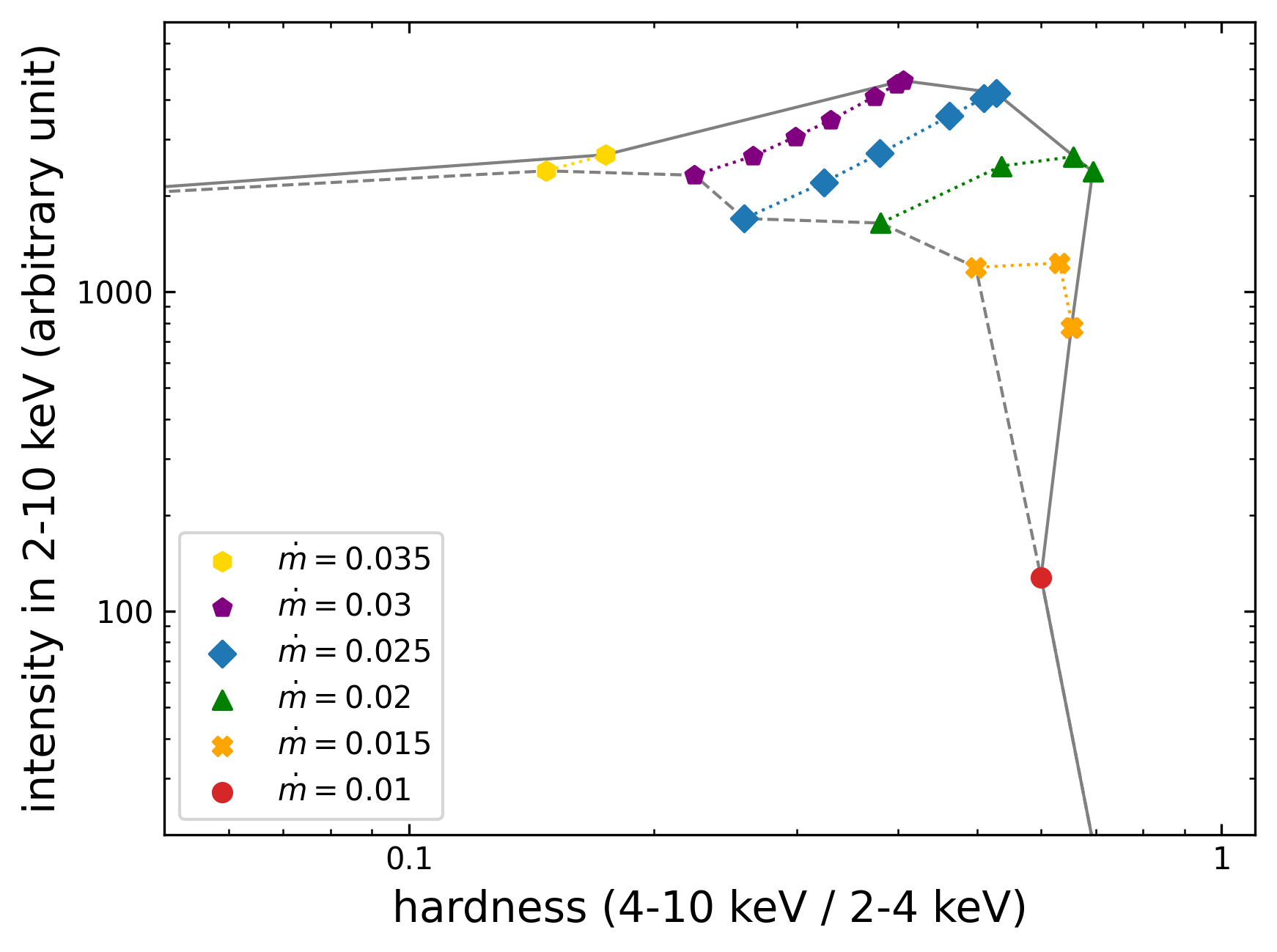}
\caption{The predicted hardness-intensity diagram with hardness defined as the ratio of photon counts between 4-10 keV and 2-4 keV bands and intensity (in arbitrary unit) the sum of the two bands. Different symbols correspond to different total accretion rate, $\dot{m}$, and different points with the same symbol have different initial corona accretion rate, $\dot{m}_{\rm cor,ini}$. The solid line marks an 'upper track' where $\dot{m}_{\rm cor,ini}$ always takes the highest possible value (see Fig. \ref{fig:mc_max}). The dashed line instead marks a 'lower track' where $\dot{m}_{\rm cor,ini}$ takes the lowest value allowed by evaporation beyond $R_{\rm out} = 500 R_{\rm S}$, which can be as high as $\sim 0.01$ \citep[e.g.][]{meyer2000b,liu2002,liu2022}. }
\label{fig:HID}
\end{figure}

We explore how the observed spectrum of BHXRBs depends on the overall accretion rate ($\dot{m}$) and the composition of the gas supply (through $\dot{m}_{\rm cor,ini}$) at $R_{\rm out}=500 R_{\rm S}$. To achieve this, for each given value of $\dot{m}$, we take a series of possible values for $\dot{m}_{\rm cor,ini}$ and then calculate the condensation/evaporation rate and thus the structure of the two-phase accretion flow along the radial direction. The emergent spectra are then calculated from Monte Carlo simulations. The spectral results are presented as a HID, as shown in Fig. \ref{fig:HID}.

Our calculations reveal that the accretion geometry is controlled by the total gas supply rate and the initial composition of the accretion flow. At high accretion rates ($\dot{m} \gtrsim 0.035$), accretion in the inner region is almost entirely via the disc, and the corona which sandwiches the disc, if exists, is very weak. On the other hand, at intermediate accretion rates ($0.01 \lesssim \dot{m} \lesssim 0.035$), different geometries of the accretion flow are possible. Depending on the initial composition of the accreted gas, the relative strength of the two phases of the accretion flow can vary in the inner region. For low accretion rates ($\dot{m} \lesssim 0.01$), we always assume pure hot gas accretion, as earlier investigations on evaporation \citep[e.g.][]{meyer2000b,liu2002,liu2022} have predicted an evaporation rate as high as $\sim 0.01 \dot{M}_{\rm Edd}$ beyond $R_{\rm out} = 500 R_{\rm S}$, which is enough to evaporate any pre-existing cold disc at this radius. In this case, an ADAF can be preserved in the innermost region of the accretion flow.

Since most of the emission is from the inner region of the system, the observed spectrum strongly depends on the accretion geometry in this region. The diversity of accretion geometry at intermediate accretion rates therefore leads to the variation of hardness ratio for a given $\dot{m}$, as presented in Fig. \ref{fig:HID}, where we used dotted lines to mark the possible variation of spectral shape due to the change in the composition of the accreted gas. A larger fraction of hot gas supply leads to a harder spectrum, as indicated by an increase of hardness ratio in Fig. \ref{fig:HID}. On the other hand, while the initial composition of the hot and cold gas supplies influences the hardness of the spectrum, the total accretion rate ($\dot{m}$) has a major impact on the observed flux, which can be seen by the vertical shift of the curves with different $\dot{m}$ as shown in Fig. \ref{fig:HID}.

The available range of hardness ratio at a given value of $\dot{m}$ can be associated with the spectral state transitions observed in BHXRBs. When $0.01 \lesssim \dot{m} \lesssim 0.035$, the hardness ratio covers a wide range, spanning from hard to soft spectral states. However, this range becomes very narrow for higher accretion rates and the possible values of hardness ratio all correspond to soft state spectra. This implies that a transition to the soft state is inevitable as $\dot{m}$ increases to a critical value, regardless of the form of gas supply. On the other hand, for lower accretion rates, the available range of hardness ratio recedes to the hard regime, corresponding to the hard spectral state.

Besides the spectral state transitions at low and high accretion rates, the diversity of the spectra, i.e., varying from hard to soft, at intermediate accretion rates ($0.01 \lesssim \dot{m} \lesssim 0.035$) is also of interest. Such diversity of spectra can be a result of not only the change of the total accretion rate as generally believed, but also the variation of the initial composition of hot and cold gas supplies, or new hot (cold) gas joining the cold (hot) accretion flow on the way towards the central black hole. This interprets the diverse continuum spectra observed as intermediate states and the occurrence of broad iron lines in low/hard state \citepalias[see also][]{wang2024}. In the following subsections, we present detailed results of the geometries and spectra of representative intermediate states.

\subsection{Effect of cold gas joining hot flows at intermediate accretion rates} \label{subsec:mdot}
It has been demonstrated that in wind accreting systems where the gas supply is hot and diffuse, given a low accretion rate (e.g. $\dot{m} \lesssim 0.03$, depending on viscosity parameter), the hot accretion flows can partially condense and form weak inner discs while maintaining a hard spectrum, which can explain the observed broad iron lines in the hard state of several BHXRBs \citepalias[][]{wang2024}. In this subsection we discuss how such hot accretion flows are affected by extra cold gas supply.

As an example we fix the initial accretion rate via the disc at $R_{\rm out}$ to a very low value, $\dot{m}_{\rm disc,ini} = 0.001$,  then examine the accretion geometries and emergent spectra for a range of $\dot{m}$ from $0.01$ to $0.03$.

\begin{figure*}
\centering
\includegraphics[width=\textwidth]{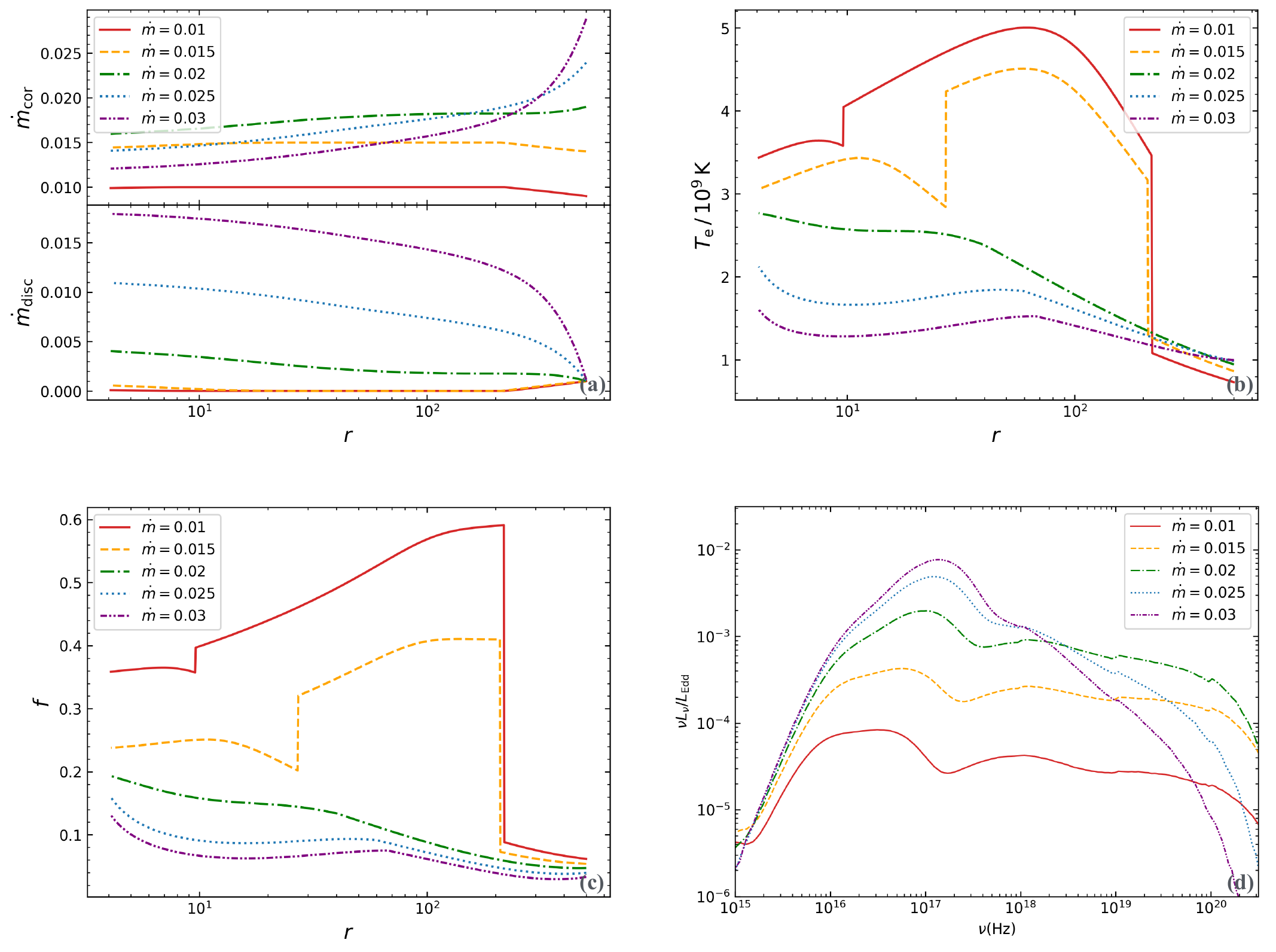}
\caption{Panel (a): the accretion rates in the corona ($\dot{m}_{\rm cor}$) and the disc ($\dot{m}_{\rm disc}$), as functions of radius. Panel (b): radial distributions of the electron temperature in the corona, $T_{\rm e}$. Panel (c): radial distributions of the advection fraction, $f$. Panel (d): the emergent spectra calculated from Monte Carlo simulations. In all panels different lines correspond to different total accretion rates, $\dot{m}$. Initial accretion rate in the disc at $R_{\rm out}$ is fixed at $\dot{m}_{\rm disc,ini} = 0.001$.}
\label{fig:fixed mdot_disc}
\end{figure*}

In the panel (a) of Fig. \ref{fig:fixed mdot_disc} we plot the accretion rates in the corona ($\dot{m}_{\rm cor}$) and the disc ($\dot{m}_{\rm disc}$) as functions of radius, $r$, which is scaled with $R_{\rm S}$. When $\dot{m}$ is lower than $\sim 0.015$, an outer evaporation region exists at $r \gtrsim 200$ where the initial weak disc is fully evaporated, and condensation occurs only in the innermost region where a weak inner disc is formed again. This result supports our basic assumption in \citetalias[][]{wang2024} that in the low/hard state (which appears when the accretion rate is low) the accretion is via an ADAF at $r \sim 100$, even in LMXBs where the accreted matter is assumed to directly feed an initial accretion disc. However, as $\dot{m}$ increases, condensation becomes more and more effective and evaporation is suppressed. When $\dot{m} \gtrsim 0.02$, evaporation disappears within the region under investigation ($r < R_{\rm out} / R_{\rm S}$) and condensation starts from the very beginning, i.e., at $R_{\rm out}$. 

The increase of $\dot{m}$ enhances condensation because gas density and hence the radiative cooling of the corona increase with $\dot{m}$, which lead to a decrease of electron temperature (see panel (b) of Fig. \ref{fig:fixed mdot_disc}) and consequently a much lower conductive flux in the transitional layer (see the strong dependence of $\Delta F_{\rm c}$ on $T_{\rm e}$ in equation (\ref{eq:new conduction})). The weak conductive heating in the transitional layer therefore results in suppressed evaporation and effective condensation. Similar conclusion has also been reached in our earlier works including \citetalias[][]{wang2024}. However, condensation in this work is much stronger as compared to \citetalias[][]{wang2024}, which is mainly the result of the introduction of an initial disc, although a lower value of $\alpha$ ($\alpha = 0.4$ was adopted in \citetalias[][]{wang2024}) and the modification on thermal conduction (see Section \ref{subsec:conduction}) also contribute to the enhancement of condensation. Even an initial disc with negligible accretion rate (e.g. $\dot{m}_{\rm disc,ini} = 0.001$ as presented here) could potentially significantly alter the property of the accretion flow, because even though a weak disc produces little radiation by itself, it can serve as a passive disc where part of the X-ray photons from coronal irradiation are reprocessed and re-emitted from the disc as thermal emission. This provides an extra source of soft photons which cool the corona by inverse Compton scattering, thus leading to earlier (in the sense that it happens at a larger radius from the black hole) and stronger condensation. Moreover, a passive disc causes thermal conduction between corona and disc, which also lowers the electron temperature in the corona and enhances condensation. The effect of an underlying disc on coronal temperature can be seen clearly from the two lines with $\dot{m} = 0.01$ and $\dot{m} = 0.015$ in panel (b) of Fig. \ref{fig:fixed mdot_disc}, where the abrupt jumps in temperature correspond to the radii at which the disc vanishes by evaporation or forms from condensation.

Panel (c) of Fig. \ref{fig:fixed mdot_disc} shows the radial distribution of the advected fraction of the viscously dissipated energy in the corona, which reflects the radiative efficiency of an accretion flow, e.g., a typical ADAF has $f \approx 1$ while a Shakura-Sunyaev disc has $f \approx 0$. Panel (c) of Fig. \ref{fig:fixed mdot_disc} shows that the radiative efficiency of the corona is higher when a disc is present, which is the result of extra Compton cooling from disc soft photons and thermal conduction.

Panel (d) of Fig. \ref{fig:fixed mdot_disc} shows the emergent spectra of the two-phase accretion flows calculated by Monte Carlo simulations. With the increase of $\dot{m}$, the soft X-ray emission from the disc continues to rise, consistent with the increase of disc accretion rate shown in panel (a) of Fig. \ref{fig:fixed mdot_disc}. On the other hand, the hard X-ray emission from the corona rises first when $\dot{m}$ increases from $0.01$ to $0.02$, but starts to decline when $\dot{m}$ gets higher. This is because condensation and evaporation processes are both inefficient and hardly change the hot accretion flow for $\dot{m} <0.02$, while condensation is enhanced significantly with increasing accretion rate for $\dot{m} >0.02$, leading to a weaker corona at higher $\dot{m}$. The radial distributions of the hot flows are displayed in the panel (a) of Fig. \ref{fig:fixed mdot_disc}, which indeed also shows that the corona accretion rate ($\dot{m}_{\rm cor}$) in the inner region which dominates the observed radiation is highest when $\dot{m} = 0.02$, and the effective condensation leaves a weaker corona near the black hole when $\dot{m} > 0.02$, even though the accretion rate in the corona at $R_{\rm out}$ is higher. Finally, panel (d) of Fig. \ref{fig:fixed mdot_disc} shows that the broadband radiation is more luminous at higher $\dot{m}$, which is the result of not only the larger amount of accreted matter, but also the increase of the overall radiative efficiency, as is reflected by the decrease of advection fraction shown in the panel (c) of Fig. \ref{fig:fixed mdot_disc}.

\subsection{Effect of the composition of the two-phase gas supply } \label{subsec:composition}
In the regime of low accretion rates ($\dot{m} \lesssim 0.01$), the inner accretion flow has been widely believed to be in the form of an ADAF. Specifically, it has been verified that in the case of pure cold gas supply, the cool disc can be completely evaporated into hot flows at a large distance from the black hole \citep[e.g.][]{meyer2000b,liu2002,liu2022}, and in the case of pure hot gas supply, the hot flow can be preserved at all radii without any condensation \citepalias[e.g.][]{wang2024}, which led to the assumption we adopted in the previous subsections that evaporation beyond $R_{\rm out} = 500 R_{\rm S}$ sets a lower limit of $\sim 0.01$ to the corona accretion rate at this radius. However, in this subsection, in order to investigate how the geometry and emergent spectrum of the inner accretion flow may vary with the composition of the gas supply, we relax the constraint on the corona accretion rate while fixing the value of the total accretion rate.   

We fix the total amount of the accreted gas at $\dot{m} = 0.01$, and take a series of values of initial disc accretion rate ranging from $0$ to $0.009$, which is equivalent to changing the initial corona accretion rate from $0.01$ to $0.001$. Since our model works only when a corona is present, we cannot provide the results with $\dot{m}_{\rm disc,ini} = \dot{m}$, instead we use $\dot{m}_{\rm disc,ini} = 0.009$ to represent the limit where the accretion is supplied almost exclusively by cold gas.

\begin{figure*}
\centering
\includegraphics[width=\textwidth]{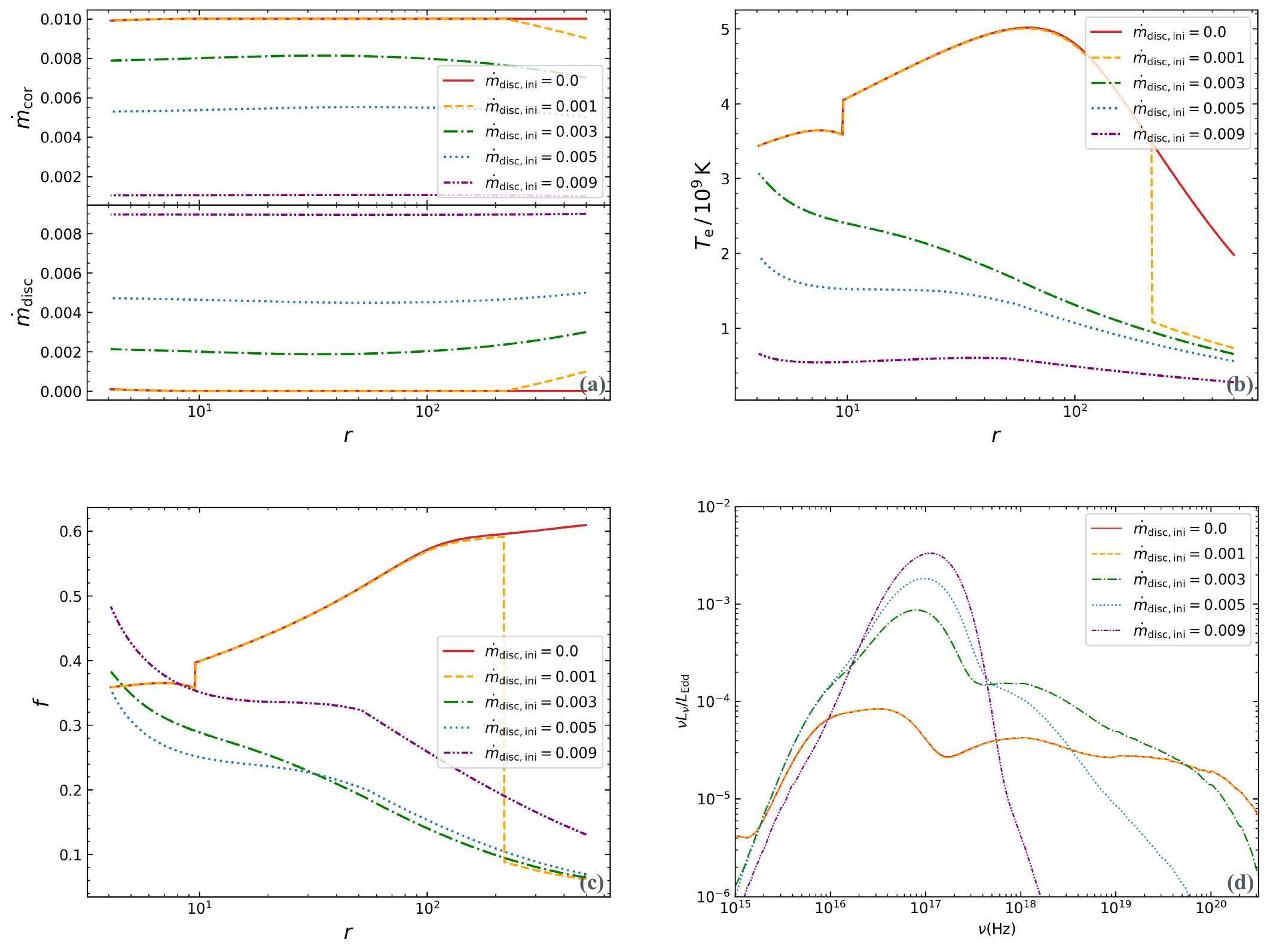}
\caption{Panel (a): the accretion rates in the corona ($\dot{m}_{\rm cor}$) and the disc ($\dot{m}_{\rm disc}$), as functions of radius. Panel (b): radial distributions of the electron temperature in the corona, $T_{\rm e}$. Panel (c): radial distributions of the advection fraction, $f$. Panel (d): the emergent spectra calculated from Monte Carlo simulations. In all panels different lines correspond to different initial disc accretion rates at $R_{\rm out}$, $\dot{m}_{\rm disc,ini}$. The total accretion rate is fixed at $\dot{m} = 0.01$.}
\label{fig:fixed mdot}
\end{figure*}

Similar to Section \ref{subsec:mdot}, we plot the radial distributions of accretion rates in corona and disc, electron temperature, advection fraction and emergent spectra respectively in panels (a)$-$(d) of Fig. \ref{fig:fixed mdot}. In each panel we see that $\dot{m}_{\rm disc,ini} = 0$ and $\dot{m}_{\rm disc,ini} = 0.001$ give basically the same results, because the weak initial disc is fully evaporated before it reaches the inner region. However, as $\dot{m}_{\rm disc,ini}$ gets higher, evaporation can no longer truncate the disc, and even though condensation happens in the inner region, it is also so weak that the accretion rates in corona and disc are essentially unchanged (see the panel (a) of Fig. \ref{fig:fixed mdot}). The suppression of evaporation is a consequence of the increase of cold flow and hence the decrease of hot flow for a given total accretion rate ($\dot m=0.01$). The former enhances external Compton cooling of the corona and the latter reduces the heating of electrons by Coulomb collisions, and thus the energy difference between heating and  cooling in the corona decreases. Therefore less energy can be used to heat up and evaporate cold gas, and for sufficiently high $\dot m_{\rm disc,ini}$, eventually a part of the coronal gas is over-cooled and condenses into the disc when the energy difference becomes negative. A consequence of this is that the electron temperature is lower for higher $\dot m_{\rm disc,ini}$, as shown in the panel (b) of Fig. \ref{fig:fixed mdot}. 

The radial distribution of the advected fraction ($f$) of the viscously dissipated energy in the corona, which reflects radiative efficiency, is shown in the panel (c) of Fig. \ref{fig:fixed mdot}. The variation of $f$ with the increase of accretion rate in the disc is caused by the two competing trends of the enhancement of radiation by more external Compton scattering and the reduction of radiation by weaker Coulomb collisional heating of electrons.     

The emergent spectra are presented in the panel (d) of Fig. \ref{fig:fixed mdot}, which shows that the spectrum becomes softer with more gas accreting via the disc, and finally turns from a hard spectrum into a soft state spectrum as cold gas supply becomes dominant. 

In this subsection we demonstrate that, at a relatively low value of total accretion rate, e.g., $\dot{m} = 0.01$, the geometries and the emergent spectra of accretion flows can be diversified for different compositions of hot and cold gas supplies. However, if $\dot{m}$ is instead fixed to a higher value, e.g., $\dot{m} = 0.05$, the disc would always be dominant regardless of the initial composition of the accretion flow. This shows that the existence of various spectra at intermediate luminosities is possible, while only soft spectra are expected at high luminosities.

Combined with the effect of cold gas joining at intermediate accretion rates of $0.01$-$0.03$ as discussed in Section \ref{subsec:mdot}, we conclude that the two-phase gas supply can result in a variety of spectra at an intermediate Eddington ratio, depending on the fraction of cold or hot gas supply, as well as the total accretion rate. This is the physical origin of the HID as shown in Fig. \ref{fig:HID}.

\subsection{Effect of the outer boundary} \label{subsec:rout}
In this work, we have set the outer radius of the accretion flow at $R_{\rm out} = 500 R_{\rm S}$, which is somewhat arbitrary. However, due to the interaction between disc and corona, the accretion geometry in the inner region and the emergent spectrum can be different if the same amount of gas ($\dot{m}$) with the same composition ($\dot{m}_{\rm disc,ini}$) is instead captured by the black hole at a different distance. Therefore, in this subsection we investigate how the change of this out boundary ($R_{\rm out}$) might affect the inner accretion flow and the spectrum. 

\begin{figure*}
\centering
\includegraphics[width=\textwidth]{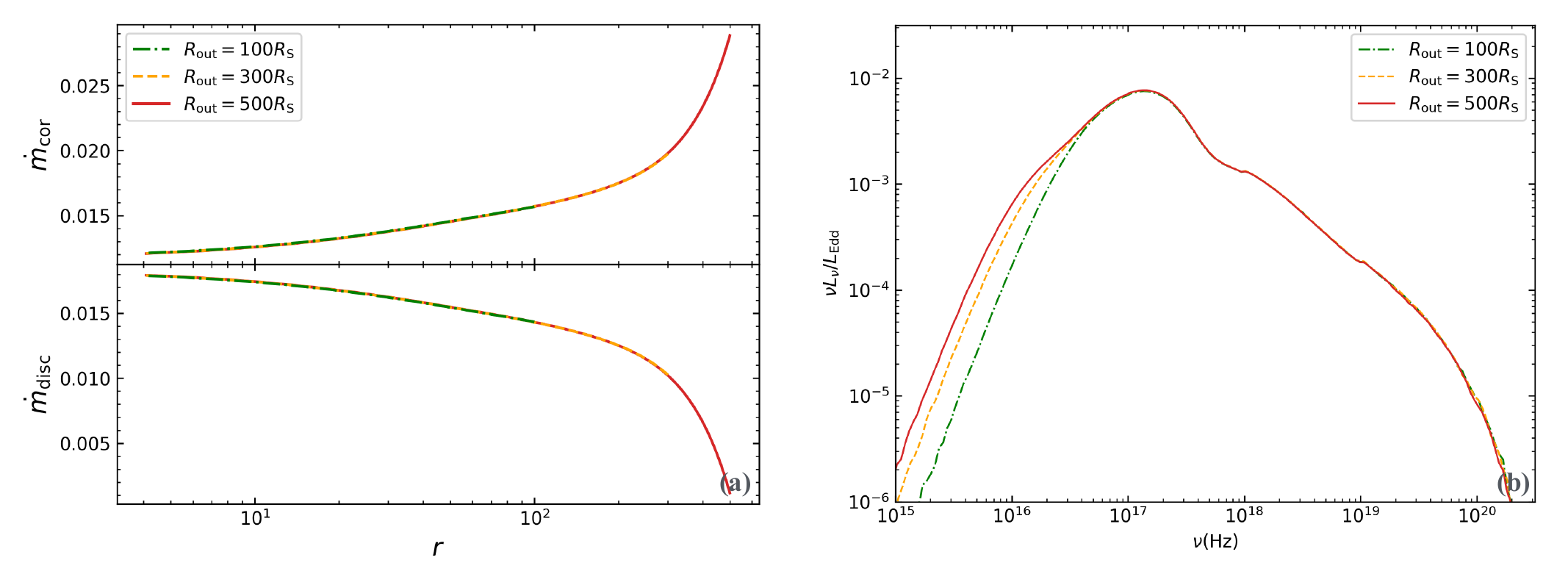}
\caption{Panel (a): the accretion rates in the corona ($\dot{m}_{\rm cor}$) and the disc ($\dot{m}_{\rm disc}$), as functions of radius. Panel (b): the emergent spectra calculated from Monte Carlo simulations. In both panels, different lines correspond to different outermost radii of the accretion flow, $R_{\rm out}$, where our calculations start. The line with $R_{\rm out} = 500 R_{\rm S}$ is calculated with $\dot{m} = 0.03$ and $\dot{m}_{\rm disc,ini} = 0.001$, and its disc accretion rates at $r = 300$ and $r = 100$ are used respectively as initial disc accretion rates ($\dot{m}_{\rm disc,ini}$) for the results with $R_{\rm out} = 300 R_{\rm S}$ and $R_{\rm out} = 100 R_{\rm S}$. }
\label{fig:rout}
\end{figure*}

As an example, we assume that the gas is supplied at $R_{\rm out} = 500 R_{\rm S}$ with $\dot{m} = 0.03$ and $\dot{m}_{\rm disc,ini} = 0.001$, for which the radial distributions of accretion rates in corona and disc have already been calculated and displayed in the panel (a) of Fig. \ref{fig:fixed mdot_disc}. We then take the values of $\dot{m}_{\rm disc}$ and $\dot{m}_{\rm cor}$ at $r = 300$ and $r = 100$ as new boundary conditions, that is, as the initial compositions of the gas supplies which are assumed to be captured respectively at $R_{\rm out} = 300 R_{\rm S}$ and $R_{\rm out} = 100 R_{\rm S}$, and perform two sets of new calculations. In the panel (a) of Fig. \ref{fig:rout} we plot the radial distributions of accretion rates. It can be seen that different choices of $R_{\rm out}$ give almost identical two-phase accretion flows in the inner region. Such results imply that the two-phase accretion flow forming at the boundary $R_{\rm out} = 500 R_{\rm S}$ with a certain composition is equivalent to that forming at another radius with an initial composition which is the same as the local composition of the accretion flow with $R_{\rm out} = 500 R_{\rm S}$ at this new radius. In other words, the results obtained with $R_{\rm out} = 500 R_{\rm S}$ as shown in Fig. \ref{fig:fixed mdot_disc} and Fig. \ref{fig:fixed mdot} can also be derived if instead the gas is supplied from another closer distance ($R_{\rm out}< 500 R_{\rm S}$), and the only difference is that the initial composition of the gas is changed to the values determined by the corresponding curves of accretion rates shown in Fig. \ref{fig:fixed mdot_disc} and Fig. \ref{fig:fixed mdot}. In addition, the above results indicate that the outer part of the accretion flow does not have a significant influence on its inner structure. The panel (b) of Fig. \ref{fig:rout} shows the emergent spectra. As expected, the X-ray emissions dominated by the inner region are the same, and the differences are in the UV band which is observationally less important for BHXRBs. 

These results demonstrate that the conclusions we made in this work are not affected by the choice of $R_{\rm out}$, so long as it is not too close to the X-ray-emitting inner region.

\section{Discussion} \label{sec:discussion}

\subsection{Cygnus X-1 and black hole high-mass X-ray binaries} \label{subsec:hmxb}
Cygnus X-1 is a prototype of HMXBs where the central black hole is believed to accrete via the stellar wind of its supergiant companion \citep[e.g.][]{lamers1976}. Because the stellar wind is hot and diffuse in space, it is naturally assumed that the gravitationally captured gas is in the form of a geometrically thick hot accretion flow. In \citetalias{wang2024} we have demonstrated that the hard state spectrum of Cygnus X-1 can be explained with pure hot gas accretion, but we were not able to increase the  accretion rate to values above the upper limit of a hot flow for interpreting the soft state. An alternative interpretation of the soft spectra with pure hot gas supply is to choose different viscous parameters respectively for the low and high states \citep[e.g.][]{taam2018, meyer-hofmeister2020} so that strong condensation can occur to form a strong disc in the inner region. In this work, we adopt two-phase accretion flows at the outer boundary when the accretion rate exceeds the upper limit of the hot flow. This is possible if the hot accretion flow (i.e. the corona) goes through an unstable phase where the extra gas collapses into a disc due to instability. By the time the accretion flow reaches equilibrium, we have a two-phase boundary condition where the accretion rate in the corona is at the maximum value, i.e., $\dot{m}_{\rm cor,ini} = \dot{m}_{\rm cor,ini,max}$, and the initial disc accretion is of course given by mass conservation as $\dot{m}_{\rm disc,ini} = \dot{m} - \dot{m}_{\rm cor,ini,max}$ (see Section \ref{subsec:constraint} and Fig. \ref{fig:mc_max}). In other words, for wind accretion systems such Cygnus X-1, we can assume that the accreted wind always feeds the corona as much as possible, and any extra gas forms an initial disc instead. The evolution of such a system in the HID there is therefore expected to follow the 'upper track' shown in Fig. \ref{fig:HID}.

\begin{figure}
\centering
\includegraphics[width=\columnwidth]{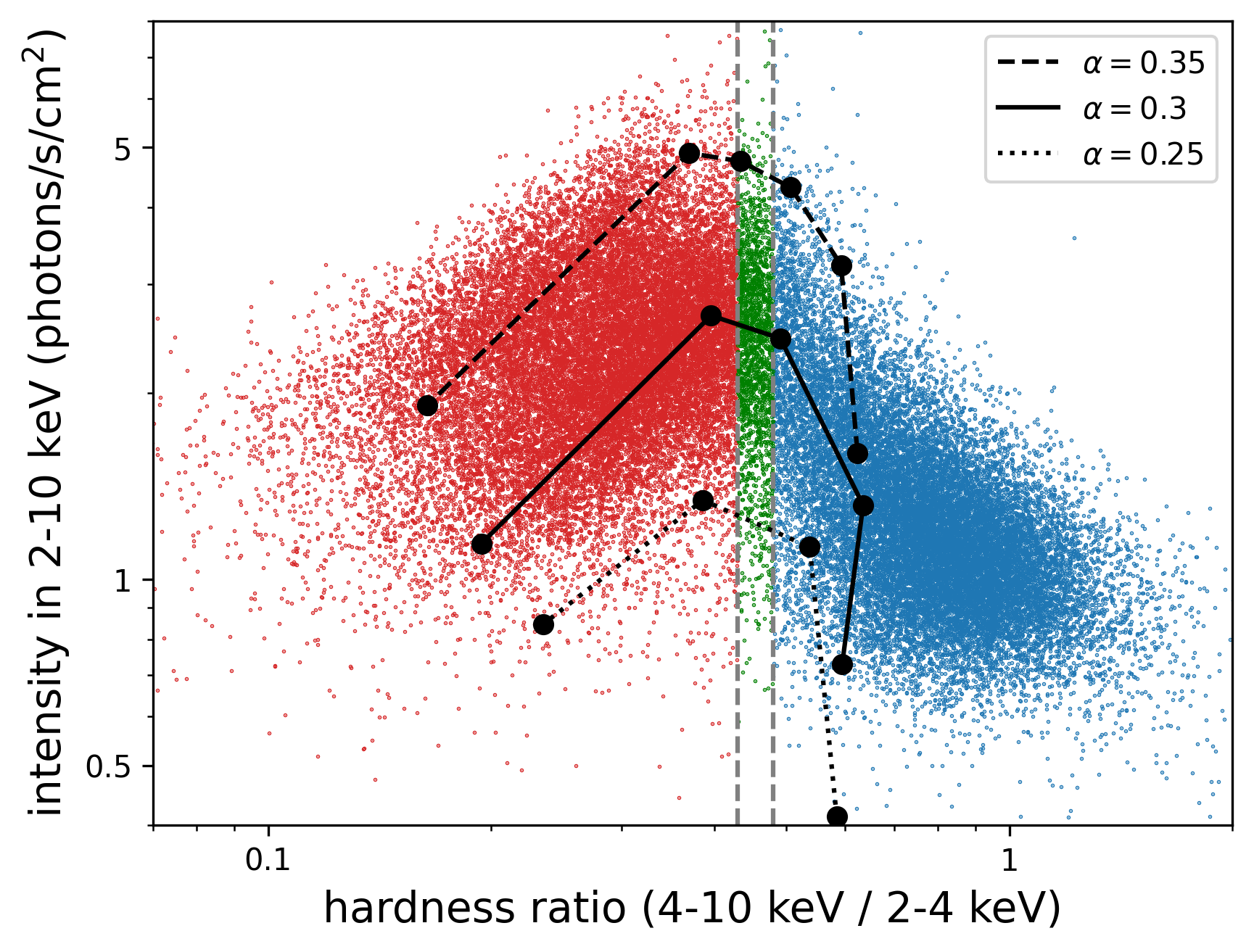}
\caption{The theoretically derived and observed hardness-intensity correlations of Cygnus X-1 in its three spectral states. The two vertical dashed lines at ${\rm hardness\ ratio} = 0.43, 0.48$ divide the observed data points into three parts, with the left part being high/soft state, the middle part intermediate state and the right part low/hard state \citep[][]{sugimoto2016}. Different curves correspond to slightly different viscosity parameters ($\alpha$). For a given $\alpha$, the theoretical results are derived by varying the total accretion rate ($\dot{m}$), while the initial accretion rate in the hot flow is assumed to always be at the upper limit, corresponding to the 'upper track' shown in Fig. \ref{fig:HID}. }
\label{fig:cyg X-1}
\end{figure}

In Fig. \ref{fig:cyg X-1} we directly compare the model predictions on the hardness-intensity correlation of Cygnus X-1 with observation. The observational data of Cygnus X-1 are obtained by the Monitor of All-sky X-ray Image (MAXI) on board the International Space Station, and we use all the available data of MAXI from August 2009 to November 2024. Following the work of \citet{sugimoto2016}, the hardness ratio is defined as the ratio of photon counts between 4-10 keV and 2-4 keV bands, and the photon counts in the 2-10 keV band (i.e. the intensity) are approximated by counts in 2-20 keV band minus those in 10-20 keV band. The observational data are classified into three spectral states with boundaries of the states defined in \citet{sugimoto2016}. The theoretical results shown in the figure are calculated with three different choices of the viscosity parameter, $\alpha = 0.25, 0.3, 0.35$. For each fixed $\alpha$, we vary the total accretion rate ($\dot{m}$) while always adopting the strongest hot flow at the outer boundary due to wind accretion, i.e., $\dot{m}_{\rm disc,ini} = \dot{m} - \dot{m}_{\rm cor,ini,max}$. Specifically, for $\alpha = 0.25$ we adopt $\dot{m} = 0.025, 0.02, 0.015, 0.012$, for $\alpha = 0.3$, $\dot{m} = 0.035, 0.03, 0.025, 0.02, 0.017$, and for $\alpha = 0.35$, $\dot{m} = 0.05, 0.045, 0.04, 0.035, 0.03, 0.025$. In the calculations of the theoretical results, we take into account the distance of Cygnus X-1, $d \approx 2.22$ kpc, its inclination of $i \approx 27.51^{\circ}$ and the updated black hole mass of $m \approx 21.2$ \citep[][]{miller-jones2021}. We note that here we also adopt an outer radius of $R_{\rm out} = 500 R_{\rm S}$, much smaller than the typical capture radius of stellar wind in such systems, which can be of order $10^5 R_{\rm S}$ \citep[e.g.][]{taam2018}{}. However, as we have demonstrated in Section \ref{subsec:rout}, this choice of $R_{\rm out}$ does not change our main conclusions. 

As is presented in Fig. \ref{fig:cyg X-1}, the main trend of the spectral behavior of Cygnus X-1 as observed by MAXI in its three spectral states can be reproduced by our model, and the viscosity parameter is constrained to a range between $0.25$ and $0.35$.

\subsection{Black hole low-mass X-ray binaries}

LMXBs are believed to accrete mainly via the Roche lobe overflow (RLOF) where the accreted gas is assumed to directly feed a disc. To explain the transition from a disc-dominated soft state to the hard state, it is conventionally predicted that the accretion disc is truncated and its inner region is replaced by a hot accretion flow as the accretion rate decreases, thus producing the observed low/hard state spectrum \citep[e.g.][]{esinetal1997, liu1999, meyer2000a, liu2005}. As for the physical mechanism behind the disc truncation, evaporation process \citep[][]{meyer1994,liufk1995,liu1999,meyer2000a,meyer2000b,rozanska2000a,rozanska2000b,liu2002,meyer-hofmeister2003,meyer-hofmeister2005} is the most promising mechanism.  

Since the RLOF in LMXBs is expected to be cold and feed the disc accretion, the initial conditions for LMXBs should be predominated by the discs, unlike the case of Cygnus X-1. However, as we have discussed in Section \ref{subsec:HID}, since the true outer boundaries of RLOFs can be much larger than the value of $R_{\rm out} = 500 R_{\rm S}$ adopted in this work where our calculations start, it is possible that part of the cold gas is evaporated beyond $R_{\rm out} = 500 R_{\rm S}$ and an initial corona exists at this radius. For simplicity we have assumed that evaporation beyond $R_{\rm out} = 500 R_{\rm S}$ gives a lower limit of up to $\sim 0.01$ to the initial corona accretion rate, according to the earlier works of \citet[][]{meyer2000b,liu2002,liu2022}. The evolution of a LMXB in the HID during its outburst then is expected to follow the 'lower track' as shown in Fig. \ref{fig:HID}. Similar to the case of Cygnus X-1, we also attempted to use the model results to interpret the observed hardness-intensity correlation of GX 339-4, a typical LMXB. We find that the predicted range of hardness ratio is consistent with observation. However, the model prediction on the highest intensity in the hard state is too low as compared to that observed in GX 339-4, although it is in agreement with the state transition in Cygnus X-1. The theoretically derived intensity could of course be increased if we adopt a higher value of $\alpha$, as is shown in Fig. \ref{fig:cyg X-1}, but to match the intensity observed in the most luminous hard state of GX 339-4, it would require an unreasonably high value of $\alpha$. To truly resolve this problem, it is necessary to introduce other heating mechanisms in the corona, which could counter the effective radiative cooling at high accretion rates, allowing a strong corona to exist and hence producing the observed luminous hard state. A candidate for the extra heating mechanism could be magnetic reconnection, which has been used to explain the X-ray emissions in AGNs \citep[e.g.][]{liu2003, liujy2024}.

\section{Conclusions} \label{sec:conclusions}
In this work, we investigate the geometry and the resultant emergent spectrum of accretion flows in BHXRBs. We make the assumption that the accreted matter is a mixture of hot and cold gases, where the hot, diffuse gas supplies a corona and the cold gas supplies a disc. Then the accreted gas forms a sandwich disc-corona geometry, and the interactions between corona and disc are investigated in detail, which include energy coupling and matter exchange, i.e., corona condensation and disc evaporation. This is our first work to incorporate condensation and evaporation processes in one single model. 

We find that the equilibrium geometry of an accretion flow and its emergent spectrum depend strongly on the amount and the initial composition of the accreted gas. More specifically, when the accretion rate is low ($\dot{m} \lesssim 0.01$), the inner accretion flow is always in the form of an ADAF. When the accretion rate is intermediate ($0.01 \lesssim \dot{m} \lesssim 0.035$), the geometry can be diversified for different initial compositions and different accretion rates. Generally speaking, in this accretion rate range, if the hot gas dominates initially, accretion in the inner region is dominated by the corona, which leads to harder spectrum, and on contrary, accretion is mainly via the disc if more cold gas is accreted. When the accretion rate is higher ($\dot{m} \gtrsim 0.035$), the cooling and ion-electron Coulomb coupling in the corona would become so effective that the corona easily condenses into the disc, leaving only a negligible corona. Therefore accretion is dominated by the disc, regardless of the form of the accreted gas. Our results interpret the spectral transitions between low and high states, as well as the various spectra, i.e., hard, intermediate and soft, observed in the intermediate states. 

As an example, we apply the model to Cygnus X-1, a wind-fed HMXB, and demonstrate that the predicted hardness-intensity correlation of Cygnus X-1 is in agreement with that observed by MAXI in its three spectral states. In addition, the viscosity parameter is constrained to the range of $0.25$--$0.35$. For LMXBs such as GX 339-4, as compared to observation, the model predicts similar spectral shapes (hardness ratios), but with much lower intensities, which hints at the necessity of introducing new heating mechanisms for the corona.

\section*{Acknowledgements}
This work is supported by the National Natural Science Foundation of China (grants 12333004 and 12433005).

%%%%%%%%%%%%%%%%%%%%%%%%%%%%%%%%%%%%%%%%%%%%%%%%%%
\section*{Data Availability}
The code used in this work will be shared on reasonable request to the corresponding authors. The MAXI data of Cygnus X-1 are publicly available at \href{http://maxi.riken.jp/star_data/J1958+352/J1958+352.html}{the MAXI home page}.

%%%%%%%%%%%%%%%%%%%% REFERENCES %%%%%%%%%%%%%%%%%%

% The best way to enter references is to use BibTeX:

\bibliographystyle{mnras}
\bibliography{accretion} % if your bibtex file is called example.bib

% Alternatively you could enter them by hand, like this:
% This method is tedious and prone to error if you have lots of references
%\begin{thebibliography}{99}
%\bibitem[\protect\citeauthoryear{Author}{2012}]{Author2012}
%Author A.~N., 2013, Journal of Improbable Astronomy, 1, 1
%\bibitem[\protect\citeauthoryear{Others}{2013}]{Others2013}
%Others S., 2012, Journal of Interesting Stuff, 17, 198
%\end{thebibliography}

%%%%%%%%%%%%%%%%%%%%%%%%%%%%%%%%%%%%%%%%%%%%%%%%%%

%%%%%%%%%%%%%%%%% APPENDICES %%%%%%%%%%%%%%%%%%%%%

%\appendix

%\section{Some extra material}

%%%%%%%%%%%%%%%%%%%%%%%%%%%%%%%%%%%%%%%%%%%%%%%%%%

% Don't change these lines
\bsp	% typesetting comment
\label{lastpage}
\end{document}